\documentclass[aps,nofootinbib,preprintnumbers,preprint]{revtex4-1}
\usepackage{geometry}     
\usepackage{graphicx}
\usepackage{amsmath,amssymb}  
\usepackage{bm}  
\usepackage{slashed}
\usepackage{hyperref}
\newcommand\beq{\begin{equation}}
\newcommand\eeq{\end{equation}}
\newcommand\Tr{\text{Tr}\,}
\newcommand\nn{\nonumber}
\newcommand\vev[1]{\langle #1\rangle}

\newcommand\hc{\text{h.c.}}

\begin{document}
\title{A Very  Light Dilaton}

\author{Benjam\'\i{}n Grinstein}
\email[Electronic address: ]{bgrinstein@ucsd.edu}
\affiliation{Department of Physics, University of California at San Diego, La Jolla, CA 92093}

\author{Patipan Uttayarat}
\email[Electronic address: ]{puttayarat@physics.ucsd.edu}
\affiliation{Department of Physics, University of California at San Diego, La Jolla, CA 92093}

\preprint{UCSD PTH 11-07}

\begin{abstract}
  We present a completely perturbative model that displays behavior similar to that of walking technicolor. In one phase of the model RG-trajectories run towards an IR-fixed point but approximate scale invariance is spontaneously broken before reaching the fixed point. The trajectories then run away from it and a light dilaton appears in the spectrum. The mass of the dilaton is controlled by the ``distance'' of the theory to the critical surface, and can be adjusted to be arbitrarily small without turning off the interactions. There is a second phase with no spontaneous symmetry breaking and hence no dilaton, and in which RG trajectories do terminate at the IR-fixed point.
\end{abstract}


\maketitle

\section{Introduction}
The Nambu-Goldstone boson of spontaneously broken scale invariance is
known as a dilaton. The name is also used to describe the pseudo
Nambu-Goldstone boson, a massive state that appears when scale
invariance is slightly broken. Classically this notion makes good
sense. For example, take a scale invariant field theory, one with only
dimensionless couplings,\footnote{In this work we consider only field
  theories in four space-time dimensions.}  with a flat direction for
the minima of the potential for scalar fields. A dilaton follows from
expanding about a non-zero field value.  Adding arbitrarily small
terms with dimensional couplings will generally give the dilaton a
small mass. However, ordinarily the passage to the quantum case can
destroy this picture. Quantum effects break scale invariance even in
the absence of explicit mass terms. The state that before quantization
would have been identified as a dilaton acquires a mass that is not
small. In fact, it is not clear one can uniquely identify a state with
what would have been the dilaton. What is meant by a ``small'' mass is
that it can be made arbitrarily small while keeping all the remaining
spectrum roughly constant and interacting. However it is not easy to
construct models displaying this behavior, that is, models of a very
light dilaton.

In their celebrated analysis of the massless abelian $U(1)$ model
Coleman and Weinberg find a scalar of mass $m$ and a vector of mass
$M$ in the spectrum, with $m^2/M^2=3e^2/8\pi^2$ ~\cite{Coleman:1973jx}.
Since the model is classically scale invariant one is tempted to
identify the only scalar with the pseudo Nambu-Goldstone boson of
broken scale invariance. It is not clear that this identification
makes sense. But even if we insist on it we see that the dilaton can
only be made light by turning off the interactions,
$e^2\to0$. Moreover, if we insist in keeping the scale of symmetry
breaking fixed then in this limit the vector meson mass also
approaches zero, albeit at a slower rate.

One may guess that a good search strategy for a light dilaton model
is to take as a starting point an exactly conformal model. Then look
to spontaneously break scale invariance and finally add small explicit
scale symmetry breaking terms. But this strategy has proven
ineffective. Consider, for example, $N=4$ supersymmetric Yang-Mills
theory, an exactly conformal interacting theory. The scalar potential
has minimum energy flat directions and one can choose to expand about a
non-trivial vacuum. Scale invariance is spontaneously
broken and a massless dilaton must emerge. However, supersymmetry is
not broken and a lot more massless stuff emerges too. As the vacuum
breaks the Yang-Mills symmetry group $G$ to one of its maximal
subgroups $H$ a full $N=4$ $H$-gauge theory remains in the massless
spectrum. The potential again has many zero energy flat directions and
we are free to identify these with ``dilatons.'' Of course, we could
just as well have identified with dilatons the flat directions of the
original theory, based on $G$. Moreover, adding perturbations will
render the dilaton very heavy, calling into question the
identification of any one state with the dilaton. A perturbation,
either relevant or  marginal, vitiates the cancellations that give
vanishing beta functions and the theory runs to strong coupling in the
infrared.   

In this work we construct a model of a light dilaton. The strategy,
construction of the model and the results of our analysis are easily
summarized. We look for a light dilaton in an interacting field theory
that displays a perturbative attractive infrared fixed point and
contains scalars. The idea is to look for spontaneous symmetry
breaking along a renormalization group trajectory headed towards the
fixed point.  For a specific model we take that of Banks and
Zaks~\cite{Banks:1981nn, *Belavin:1974gu} supplemented with scalars that are neutral
under the gauge group. The scalars have quartic self-interactions and
are Yukawa-coupled to the Banks-Zaks spinors. As the Yang-Mills gauge
coupling runs toward the Banks-Zaks IR-fixed point, it drives the
scalar and Yukawa couplings towards the non-trivial fixed point values
too. Depending on the relative values of the coupling constants the
Coleman-Weinberg effective potential for the scalar fields may develop
a non-trivial minimum~\cite{Coleman:1973jx}. The parameter space of the
theory is split according to whether scaling symmetry is spontaneously
broken or not, and for couplings near the boundary between these
regions the dilaton is very light in units of its decay constant. Yet
the theory is fully interacting and the spectrum is non-trivial (and
insensitive to the parameter adjustment required to make the dilaton
arbitrarily light).

Our search for a model of a very light dilaton was partially motivated
by recent work of Appelquist and Bai~\cite{Appelquist:2010gy}
(henceforth `AB') and by Hashimoto and Yamawaki
~\cite{Hashimoto:2010nw} rekindling and
old debate on whether walking technicolor (WTC) may have a light
dilaton in its spectrum~\cite{Bardeen:1985sm,  *Clark:1986gx,*Leung:1989hw,
  Yamawaki:1985zg,  *Bando:1986bg,* Nonoyama:1989dq,
  Holdom:1986ub, *Holdom:1987yu}.
The idea of ``walking'' promises to solve many difficulties of
technicolor (TC) theories. The conjectural behavior of the theory
requires that (1) the TC coupling constant $g$ evolves very slowly,
(2) this occurs while at large value of the TC coupling constant, so
that anomalous dimensions are large, and (3) the slowly running
coupling eventually crosses a threshold, exceeding a critical value
$g_c$ for chiral symmetry breaking. The picture is that once the
coupling crosses this threshold, techniquarks become massive, decouple
and leave the technigluons to drive alone the running of the coupling
constant (which from that point on grows quickly, much like in
QCD). The condensate that results breaks electroweak symmetry giving
masses to $W$ and $Z$ gauge bosons. The large anomalous dimension of
the techniquark bi-linear insures that four-fermion operators induced
by extended-TC interactions (ETC) give acceptable masses to all but
the top quarks (and leptons) while effectively suppressing ETC
mediated FCNCs.  Moreover, the large anomalous dimensions of
4-techniquark operators also induced by the ETC tend to increase the
masses of troublesome pseudo-Goldstone to acceptable levels.  In this
picture the slow evolution of the coupling constant can be viewed as
an approach towards a would-be conformal fixed point, $g_*$. It is a
``would-be'' fixed point only because $g_c< g_*$, which triggers the
fast QCD-like evolution of $g$ once it exceeds the critical value
$g_c$. AB argue, while Hashimoto and Yamawaki rebut,
that a dilaton does appear and estimate that its mass is roughly determined by
the value of the beta function at its closest approach to the would-be
fixed point,  $\beta(g_c)$.

The existence of a light dilaton in WTC is by no means
obvious. The dilaton is in some respects similar to the $\eta'$ in
QCD. Were we to ignore the $U(1)_A$ anomaly it would be a pseudo
Nambu-Goldstone boson, on par with the $(\pi, K, \eta)$ octet. But
the anomaly breaks the symmetry explicitly and because it involves
the strong interactions  this breaking is not a small perturbation. 
Beyond deciding whether the light dilaton appears in the spectrum
of WTC, there are many other questions that arise. For example, what
precisely is the meaning of the critical coupling $g_c$, what is the
dilaton decay constant, etc.

Unfortunately, as of this writing there is no explicit realization of
the WTC idea as a specific model. Numerous numerical studies are
ongoing to determine whether QCD-like theories at the edge of the
conformal window display the phenomenon of
walking~\cite{Fodor:2011tw,*Fodor:2011tu,Deuzeman:2009mh,Hasenfratz:2010fi,Catterall:2009sb,Hietanen:2008mr,*Hietanen:2009az,Bursa:2011ru,*Bursa:2009we,*Bursa:2010xn,Andersen:2011yj,DelDebbio:2010hu,*DelDebbio:2010hx,DeGrand:2011qd,*DeGrand:2010na,*DeGrand:2008kx,Kogut:2010cz,*Sinclair:2010be,Hayakawa:2010yn,*Yamada:2010wd,Ohki:2010sr,*Jin:2009mc,*Bilgici:2009nm}.
While a positive result from these studies may confirm the existence of
models exhibiting the WTC idea, a negative result would not rule out
the possibility that some non-QCD like theory behaves this way. In the
mean time it would be useful to construct a toy model displaying some
of the WTC behavior. One would like the toy model to be fully
perturbative so that one may readily compute and resolve questions. In
some ways our model fits the bill. It does have coupling constants
that grow as they approach a fixed point, then walk for quite a long
RG-time and finally swerve away. This change of behavior is triggered,
much like in the WTC idea, by the analogue of chiral symmetry
breaking, that is, the scalar fields acquiring a non-trivial
expectation value, giving masses to the spinors through their Yukawa
couplings. To be sure, the model fails to mimic WTC in important
ways. By design it remains perturbative, and therefore anomalous
dimensions remain small. And, as opposed to a would be WTC theory, our
model is not asymptotically free; while the Banks-Zaks sector is,
RG-running in the scalar sector encounters Landau poles. We do not see
the latter of these difficulties as central. One can view this as a
theory with a cut-off at a scale that is exponentially large compared
to where the physics of the symmetry breaking takes place, or imagine
that it is the low energy effective theory of a more complete model.

But the usefulness of an explicit model of a very light dilaton goes
beyond that of being a toy WTC. Sundrum has remarked that the dilaton
can serve as a scalar analog of the graviton.  By studying the
properties of the dilaton one can hope to gain insights into the
theory of gravity and perhaps find the answer to the cosmological
constant puzzle~\cite{Sundrum:2003yt}. A dilaton is also likely to
appear in the AdS/CFT dual of the Randal-Sundrum
model~\cite{Randall:1999ee} with the Goldberger-Wise mechanism
stabilizing the extra-dimension~\cite{Goldberger:1999uk}. In the
4-dimensional language, the theory is described not by a CFT but by a
flow to a CFT fixed point which is however interrupted close to the
fixed point by the expectation value of a field that measures the
distance from the origin in moduli
space~\cite{ArkaniHamed:2000ds,Rattazzi:2000hs}. This is described
effectively by a theory at the fixed point, a CFT Lagrangian,
supplemented by small perturbations. The latter are made scale
invariant by including couplings to the dilaton in the spirit of
phenomenological Lagrangians~\cite{Coleman:1969sm,*Callan:1969sn}. If
the SM is embedded in such a scheme the dilaton may behave much like,
but not exactly the same as, the higgs boson of the minimal standard
model~\cite{Goldberger:2008zz,Fan:2008jk,Vecchi:2010gj}. An amusing
question that one can now ponder is the inverse AdS/CFT problem: given
our perturbative model, what is the AdS dual (presumably a strongly
interacting non-factorizable gravity model in 5 dimensions)?

Another area where the dilaton may play a role is in astrophysics and
cosmology. By noting that the dilaton couples to the trace of the
stress energy tensor, the authors in Ref.~\cite{Bai:2009ms} propose to
use a light dilaton as a force mediator between the SM particles and
dark matter particles.  Some authors also propose a light dilaton as a
new dark matter candidate~\cite{Choi:2011fy}. In all these cases an
explicit computable model may be put to use in understanding issues
currently clouded by our inability to compute at or near strongly
interacting fixed points.

The paper is organized as follows.  In section 2 we introduce our
model and show the existent of both the IR-fixed point and the
non-trivial vacuum.  In section 3 we identify the state corresponding
to the dilaton and we compute its mass. In section 4 we discuss a
phase structure of our model accessible in perturbation theory. We
discuss our results briefly in Sec.~5.

\section{The Model}
We study a class of SU$(N)$ gauge theories with $n_f=n_\chi+n_\psi=2n_\chi$ flavors of
spinors, $\psi_i$ and $\chi_k$,  and two real scalars.  The spinors are taken to be vector-like
in the fundamental representation of the gauge group while the scalars
are singlets. The most general Lagrangian that is classically scale
invariant and also invariant under  the
discrete symmetry  $\phi_1\to\phi_1$, $\phi_2\to-\phi_2$, $\psi\to\psi$,
$\chi\to-\chi$,  and the global
simultaneous $SU(n_\chi)$ transformations $\psi\to U \psi$, $\chi\to U \chi$
 is 
\begin{equation}
\label{eq:model}
\begin{split}
  \mathcal{L} &= -\frac{1}{2}\Tr F^{\mu\nu}F_{\mu\nu} + \sum_{j=1}^{n_\chi} \bar{\psi}^{j}i\slashed{D}\psi_{j}+ \sum_{k=1}^{n_\chi} \bar{\chi}^{k}i\slashed{D}\chi_{k} +\frac{1}{2}(\partial_\mu\phi_1)^2+\frac{1}{2}(\partial_\mu\phi_2)^2\\
  & \quad- y_1\left(\bar{\psi}\psi+\bar{\chi}\chi
  \right)\phi_1-y_2(\bar{\psi}\chi+\hc)\phi_2 -
  \frac{1}{24}\lambda_1\phi_1^4- \frac{1}{24}\lambda_2\phi_2^4-
  \frac{1}{4}\lambda_3\phi_1^2\phi_2^2\,.
\end{split}
\end{equation}
Quantum effects will induce scalar masses of the order of the
cut-off. In the spirit of Coleman and Weinberg we happily subtract
these masses away~\cite{Coleman:1973jx}; after all, we are not
interested in solving the hierarchy problem. Alternatively one can
study this theory perturbatively in the continuum, using dimensional
regularization.

For small number of families this model is very
similar to QCD. The gauge sector will run to strong coupling in the
infrared, the remaining parameters will only act  as small
perturbations. The chiral symmetry $SU(n_f)\times SU(n_f)$ is
spontaneously  broken  to its diagonal subgroup with associated Nambu-Goldstone
bosons in the spectrum. 

We are interested in larger values of $n_f$ for which the gauge
coupling is still asymptotically free but behaves very differently in
the infrared, as we now discuss.

\subsection{Fixed Point Structure}

We arrange the values of $N$ and $n_f$ so that the coefficient in the
one-loop term of the gauge beta function is small, much as Banks and
Zaks do for QCD~\cite{Banks:1981nn}. The perturbative fixed point
value in the gauge coupling appears from balancing the one and two
loop terms against each other.  To arrange for an arbitrarily small
fixed point value we consider only large values of $N$ and $n_f$. The
coefficients of the one-loop terms of the beta functions for the
remaining couplings are not small. Hence it suffices to retain only up
to one loop order in the beta functions of Yukawa and scalar
couplings, while, of course, retaining up to two loop order for that
of the gauge coupling.  The mass independent ({\it e.g.,} minimal
subtraction) $\beta$-functions at large $N$ and $n_f$ are given
by~\cite{Jack:1982sr,*Machacek:1983tz,*Machacek:1983fi,*Machacek:1984zw}
\begin{equation}
\label{eq:betafunctions}
\begin{aligned}
  (16\pi^2)\frac{\partial g}{\partial t} &= -\frac{\delta N}{3}g^3 + \frac{25N^2}{2}\frac{g^5}{16\pi^2}\\
  (16\pi^2)\frac{\partial y_1}{\partial t} &= 4y_1y_2^2 + 11N^2 y_1^3 - 3Ng^2y_1,\\
  (16\pi^2)\frac{\partial y_2}{\partial t} &= 3y_1^2y_2 + 11N^2 y_2^3 - 3Ng^2y_2\\
  (16\pi^2)\frac{\partial\lambda_1}{\partial t} &= 3\lambda_1^2+3\lambda_3^2+44N^2\lambda_1 y_1^2 - 264N^2 y_1^4,\\
  (16\pi^2)\frac{\partial\lambda_2}{\partial t} &= 3\lambda_2^2+3\lambda_3^2+44N^2\lambda_2 y_2^2- 264N^2 y_2^4,\\
  (16\pi^2)\frac{\partial\lambda_3}{\partial t} &=
  \lambda_1\lambda_3+\lambda_2\lambda_3+4\lambda_3^2+22N^2\lambda_3
  y_1^2+22N^2\lambda_3y_2^2 -264N^2 y_1^2y_2^2.
\end{aligned}
\end{equation}
The number of families is taken to be fixed at
$n_f=11N/2(1-\delta/11)$ and we drop the $\mathcal{O}(\delta)$ terms except in $\beta_g$.
Even though $N$ and $n_f$ are integers, one can make   $\delta$  arbitrarily
small by taking  $N$ and $n_f$   arbitrarily large. 

These equations will play an important role in our discussion. The
first step is to determine whether any non-trivial fixed points
exist. To see that one does indeed run into the fix point we can argue as
follows. First, there is no question that the gauge coupling flows in
the IR towards it fixed point. All that is required is that it starts
its flow from the UV at a value smaller than the fixed point. Then the
Yukawa couplings' beta functions are dominated by the last term, which
is negative and only linear in the $y_i$'s. Hence they grow until the
positive non-linear terms compensate against the last negative, linear
term. And the story is then repeated for the scalars, but now having
the Yukawa couplings drive the beta functions (the last terms in each
of the three scalar coupling beta functions are negative and $\lambda_i$
independent).  

The mechanism that is driving the couplings towards theIR-fixed point
values is mimicked by the process of determining their location. The
gauge coupling has the same fixed point as in the Banks-Zaks
model. This is used in the equations for the Yukawa couplings
$y_{1,2}$ which are then trivially solved to leading order in $1/N$
accuracy. In turn these solutions are used in the equations for the
scalar self-couplings.  To leading order in $1/N$ accuracy, the fixed
point is at the following zeroes of the beta functions:
\begin{equation}
\begin{gathered}
	g_\ast^2 = 16\pi^2\frac{2}{75}\frac{\delta}{N}\,,\qquad
	y_{1\ast}^2 = y_{2\ast}^2= \frac{3}{11}\frac{g_\ast^2}{N}\,, \qquad
	\lambda_{1\ast} =\lambda_{2\ast} = \lambda_{3\ast} = 6y_{1\ast}^2  = \frac{18}{11}\frac{g_\ast^2}{N}.
\end{gathered}
\end{equation}
Since $\delta$ is arbitrarily small while $N$ is
arbitrarily large the fixed point values of the couplings are all
perturbative. It is easy to check that the terms omitted in the loop
expansion of the beta functions are parametrically smaller.

This result may be surprising. Common lore, which of course cannot be
documented, is that theories with scalars and fermions do not exhibit
nontrivial IR-fixed points in 4 dimensions. While this is obviously
false in $4-\epsilon$ dimensions, we see that it is also
false in exactly four dimensions. The lore's intuition is vitiated
here because it is the gauge coupling which is driving the remaining
couplings toward the fixed point.

\subsection{Vacuum Structure}
\label{sec:vac}
We turn now to the physical content of our model. The first order of
business is to understand its vacuum structure and determine the fate
of the  symmetries
of the  Lagrangian.
At the classical level, the potential is trivially minimized,
$\vev{\phi_1}=\vev{\phi_2} = 0$ and all symmetries are explicitly
realized.  However, this may change once quantum effects are
included.   The  one-loop  effective potential in the $\overline{\text{MS}}$
scheme is~\cite{Coleman:1973jx}
\begin{equation}
\label{eq:fulleffectivepotential}
\begin{split}
  V_{\text{eff}} &= - \frac{1}{24}\lambda_1\phi_1^4- \frac{1}{24}\lambda_2\phi_2^4- \frac{1}{4}\lambda_3\phi_1^2\phi_2^2\\
  &\qquad - \frac{11N^2 M_{f+}^4}{(64\pi^2)}\left(\ln\frac{M_{f+}^2}{2\mu^2}-\frac{3}{2}\right) -\frac{11N^2 M_{f-}^4}{(64\pi^2)}\left(\ln\frac{M_{f-}^2}{2\mu^2}-\frac{3}{2}\right) \\
  &\qquad
  +\frac{M_{s+}^4}{(64\pi^2)}\left(\ln\frac{M_{s+}^2}{\mu^2}-\frac{3}{2}\right)
  +
  \frac{M_{s-}^4}{(64\pi^2)}\left(\ln\frac{M_{s-}^2}{2\mu^2}-\frac{3}{2}\right),
\end{split}
\end{equation}
where
\begin{equation}
\begin{split}
  M_{f\pm} &=	y_1\phi_1 \pm y_2\phi_2 ,\\
  M_{s\pm}^2 &= \frac{(\lambda_1+\lambda_3)\phi_1^2+(\lambda_2+\lambda_3)\phi_2^2}{4}\\
  &\qquad\pm\frac{\sqrt{(\lambda_1-\lambda_3)^2\phi_1^4+(\lambda_2-\lambda_3)^2\phi_2^4-2(\lambda_1\lambda_2-\lambda_1\lambda_3-\lambda_2\lambda_3-7\lambda_3^2)\phi_1^2\phi_2^2}}{4}.
\end{split}
\end{equation}  
No mass terms have appeared because we have used dimensional
regularization (in the $\overline{\text{MS}}$ scheme).  As explained
earlier, this is in keeping with Coleman and Weinberg who completely
subtract the mass terms. We will return to this point in the
discussion where we will argue that including small masses for the
scalars and spinors of the model does not modify the main conclusions
(but we have to wait until then to explain the meaning of ``small.'')

It is fairly difficult to search for the minimum  of this
function. We can however find some local minima easily, by searching
only for a vacuum that preserves the discrete
symmetry $\phi_1\to\phi_1$, $\phi_2\to-\phi_2$, $\psi\to\psi$ and
$\chi\to-\chi$.  The effective potential along the 
$\phi_2=0$ axis is much simplified:
\begin{equation}
\label{eq:effectivepotential}
\begin{split}
  V_{\text{eff}} &= \frac{\lambda_1}{24}\phi_1^4 + \frac{(\lambda_1\phi_1^2)^2}{256\pi^2}\left(\ln\frac{\lambda_1\phi_1^2}{2\mu^2}-\frac{3}{2}\right)+\frac{(\lambda_3\phi_1^2)^2}{256\pi^2}\left(\ln\frac{\lambda_3\phi_1^2}{2\mu^2}-\frac{3}{2}\right)\\
  &\phantom{=}\quad-\frac{22N^2
    y_1^4\phi_1^4}{64\pi^2}\left(\ln\frac{y_1^2\phi_1^2}{\mu^2}-\frac32\right).
\end{split}
\end{equation}
It is straightforward to find an extremum of this function,
\begin{align}
  \frac{\partial}{\partial\phi_1}V_{\text{eff}}(\vev{\phi_1}) &= 0\nn\\
  \Longrightarrow -\frac{\lambda_1}{6} &= \frac{\lambda_1^2}{64\pi^2}\left(\ln\frac{\lambda_1\vev{\phi_1}^2}{2\mu^2}-1\right)+\frac{\lambda_3^2}{64\pi^2}\left(\ln\frac{\lambda_3\vev{\phi_1}^2}{2\mu^2}-1\right)\nn\\
  &\phantom{=}\quad-\frac{88N^2
    y_1^4}{64\pi^2}\left(\ln\frac{y_1^2\vev{\phi_1}^2}{\mu^2}-1\right) \label{eq:vev}.
\end{align}
If the extremum is a minimum this equation determines the vacuum
expectation $\vev{\phi_1}$ in terms of the coupling constants of the
model. Alternatively one may eliminate one of the dimensionless
parameters of the model in favor of the dimensional vacuum expectation
value. This is the well known dimensional transmutation
procedure. Since the expectation value sets the physical scale for the
theory we adopt this approach here so in what follows the
dimensionless parameter $\lambda_1$ is understood as a function of the
couplings and the expectation value, given in
\eqref{eq:vev}. In order that the perturbative expansion of 
$V_{\text{  eff}}$ not be invalidated by large logs in higher orders we insist
that $\lambda_1/16\pi^2\ln(\vev{\phi_1}^2/\mu^2)\ll1$. Then
$\lambda_1$ is given by the last two terms in \eqref{eq:vev}, and this
condition   becomes
\begin{equation}
\label{eq:vevCondition}
\frac{\lambda_3^2-88N^2  y_1^4}{(16\pi^2)^2}\ln^2\frac{\vev{\phi_1}^2}{\mu^2} \ll 1\,.
\end{equation}
Since $\lambda_1$ has been eliminated in favor of $\vev{\phi_1}$, the
conditions that the perturbative analysis is valid are that
$\vev{\phi_1}$ satisfies \eqref{eq:vevCondition} and that
dimensionless couplings remain small. Next, we must check that the
extremum is a local minimum and that it is of lower energy than that
of the origin of field space.

We first verify that the extremum is a local minimum.  To this end we
need to check that the eigenvalues of the mass matrix are both
positive. Owing to the discrete symmetry and the fact that we are on
the $\phi_2=0$ axis, the mixed derivatives terms vanish at
$\vev{\phi_1}$,
$\frac{\partial^2}{\partial\phi_1\partial\phi_2}V_{\text{eff}}(\vev{\phi_1},0)=0$.
Hence the two eigenvalues are given by 
\begin{align}
  \frac{\partial^2}{\partial\phi_1^2}V_{\text{eff}}(\vev{\phi_1},0) &= \frac{\lambda_3^2-88N^2 y_1^4}{32\pi^2}\vev{\phi_1}^2,\\
  \frac{\partial^2}{\partial\phi_2^2}V_{\text{eff}}(\vev{\phi_1},0) &= \frac{\lambda_3}{2}\vev{\phi_1}^2-\frac{\lambda_3(\lambda_2+4\lambda_3)\vev{\phi_1}^2}{64\pi^2}\left(\ln\frac{\lambda_3\vev{\phi_1}^2}{2\mu^2}+1\right)\nn\\
  &\phantom{=}\quad-\frac{264N^2
    y_1^2y_2^2\vev{\phi1}^2}{64\pi^2}\left(\ln\frac{y_1^2\vev{\phi_1}^2}{\mu^2}-\frac{1}{3}\right).
\end{align}
The first eigenvalue is  positive provided
\begin{equation}
\label{eq:posEig}
	\varepsilon\equiv\lambda_3^2 - 88N^2 y_1^4>0.
\end{equation}
The second eigenvalue is generally positive provided we are in the
regime where  the one loop terms
are small compared to the tree level term. This is generally the case
in perturbation theory, although one could have one coupling, in this
case $\lambda_3$ be  small compared to the remaining couplings (and
indeed this is the situation for $\lambda_1$ in the region of parameter
space of interest). 

We can now check that the effective potential at $\vev{\phi_1}$ is
negative:
\begin{equation}
  \label{eq:potentialatvev}
  V_{\text{eff}}(\vev{\phi_1}) = -\frac{\lambda_3^2-88N^2 y_1^4}{512\pi^2}\vev{\phi_1}^4
= -\frac{\varepsilon}{512\pi^2}\vev{\phi_1}^4.
\end{equation}
Remarkably, the condition that this be  negative is precisely the same
as having the first eigenvalue of the mass matrix be positive, Eq.~\eqref{eq:posEig}. 

Here we take a small detour to discuss the role of $\phi_2$.  The
readers might notice that $\phi_2$ plays virtually no role in the
above analysis of the vacuum.  Moreover, we could have arranged for
the non-trivial IR fixed-point with only one scalar. This can be seen
by setting $y_2$, $\lambda_2$ and $\lambda_3$ to zero in Eq
(\ref{eq:betafunctions}) and repeating the fixed-point analysis given
above.\footnote{Similar result regarding the fixed-point in Banks-Zaks
  type theory with an extra scalar singlet has been independently
  obtained in~\cite{Rattazzi:2008pe, *Antipin:2011ny}.} This begs the
question -- what is the purpose of $\phi_2$? With only one
scalar, $\phi_1$, we can repeat the above analysis and reproduce
Eqs.~(\ref{eq:vevCondition})--(\ref{eq:potentialatvev}) with
$\lambda_3$ set to 0. Clearly, the extremum becomes the maximum and
the effective potential seems to be unbounded from below. The extra
scalar field allow us to introduce more couplings and more importantly
establish the non-trivial minimum via perturbative analysis.

Note that the conditions we have found for the non-trivial minimum of
the effective potential are not satisfied at or in the vicinity of the
IR-fixed point. But neither are the conditions for perturbative
computability. In order to determine the vacuum structure near the IR
fixed point we must re-sum the leading log expansion of the effective
potential. Equivalently we can take any point in the vicinity of the
fixed point and ask whether its RG-trajectory maps back at some large
RG-time $t$ to the region where the analysis above is valid. If that
is the case we can further ask whether it gives a non-trivial
minimum. This is the approach we adopt here. We will come back to this
issue in Sec.~\ref{sec:phase} where we will discuss the phase
structure of the model and integrate the RGEs numerically to verify
the vacuum structure near the IR-fixed point. But even without
numerical studies we can argue physically that there are points
arbitrarily close to the IR-fixed points for which the vacuum is
non-trivial and scale invariance is spontaneously broken.

Choose the parameters to satisfy \eqref{eq:posEig} and to be small at
some fixed renormalization scale $\mu_0$. One can arrange for the
allowed range of expectation values to be large, so that $ \vev{\phi_1}\ll
\mu_0$ is included, by choosing $\varepsilon$ to be as small as necessary.  The
coupling constants will run as in the mass independent scheme until
the scale $\mu$ reaches values comparable to the mass of the heaviest
particle in the model. At that point the running is modified. The
trajectory that would end at the IR-fixed point is modified before the
fixed point is reached. However this modification to the trajectory
occurs only for $\mu\lesssim \vev{\phi_1}$. That is, given a fixed
starting point $\mu_0$ we can choose to run as far as needed on the
mass-independent trajectory, far enough that it gets arbitrarily close
to the IR-fixed point; all that is required is that one starts with a
small enough value of $\vev{\phi_1}$.

We have not been able to explore fully the landscape of our effective
potential.  Other, lower minima may exist outside the $\phi_2=0$
axis. If that is the case the minimum we have found describes only a
metastable vacuum. The analysis that follows is still largely
correct. But more importantly, an analogous analysis could be applied
to the global minimum and the qualitative results will not be
different. What is important here is that the non-trivial minimum
found at one-loop spontaneously breaks the scale invariance of the
classical Lagrangian. The scale invariance is explicitly broken at
one-loop too, by a quantum mechanical anomaly. If the former effect is
dominant then we expect to see a pseudo Nambu-Goldstone boson of
spontaneously broken approximate scale invariance, while if the latter
effect is dominant no such state will be seen. So we turn in the next
section to determining the spectrum of the model.

\subsection{Particle Spectrum}
\label{sec:spectrum} 
If the theory is in the symmetric phase,
$\vev{\phi_1}=\vev{\phi_2}=0$, then all the particles are massless.
Here, we compute the spectrum in the broken phase, $\vev{\phi_1}= v,
\vev{\phi_2}=0$. We retain up to one-loop order in the computation of
the spectrum so that we may later address questions of invariance of
physical quantities under RG-evolution. This is important because on
the one hand we determine the vacuum structure far away from the IR
fixed point while on the other we are interested in the fate of scale
invariance and hence want to study the RG flow towards, and eventually
in the vicinity of, the IR-fixed point.

We first compute the fermion spectrum. For  large $N$ 
the leading contribution to the fermion self-energy is from the gauge
interaction.  We can parametrize the self-energy as
\begin{equation}
	i\Sigma(\slashed{p}) =i(A m+B\slashed{p}).
\end{equation}
We obtain, to one-loop order,
\begin{equation}
	A = \frac{g^2}{16\pi^2}\frac{N}{2}\left(-3\ln\frac{y_1^2v^2}{\mu^2}+4\right)\,,\quad
	B = 1\,,
\end{equation}
in Landau gauge.  Hence the   masses of $\chi$ and $\psi$ (poles in the
respective propagators) are
\begin{equation}
  M_{\psi}(\mu)=M_{\chi}(\mu) = y_1v\left[1-\frac{g^2}{16\pi^2}\frac{N}{2}\left(3\ln\frac{y_1^2v^2}{\mu^2}-4\right)\right]\,.\label{eq:fermionpolemass}
\end{equation}

The pole masses of the scalar fields $\phi_1$ and $\phi_2$ can be computed in
a similar manner. Schematically, to one-loop order, the mass is
\begin{equation}
	M_\phi^2 = \frac{\lambda}{2}v^2 + \Pi(\lambda v^2/2).
\end{equation}
Explicit computation yields 
\begin{align}
  M_{\phi_1}^2 &= \frac{\lambda_1v^2}{2}+\frac{3\lambda_1^2v^2}{64\pi^2}\left(\ln\frac{\lambda_1v^2}{2{\mu}^2}-\frac{5}{3}+\frac{2\pi}{3\sqrt{3}}\right)+\frac{3\lambda_3^2v^2}{64\pi^2}\left(\ln\frac{\lambda_3v^2}{2{\mu}^2}-\frac{1}{3}-\frac{2\lambda_1}{3\lambda_3}\right)\nn\\
  &\phantom{=}\quad +\frac{22N^2 y_1^2}{16\pi^2}\Bigg[y_1^2v^2-\frac{\lambda_1v^2}{12}-3\left(y_1^2v^2-\frac{\lambda_1v^2}{12}\right)\left(\ln\frac{y_1^2v^2}{{\mu}^2}\right)\nn\\
  &\hspace{0.7in}-3\int_0^1\text{d}x\left(y_1^2v^2-\frac{x(1-x)}{2}\lambda_1v^2\right)\ln\left(1-x(1-x)\frac{\lambda_1}{2y_1^2}\right)\Bigg],\label{eq:phi1polemass}\\
&= \frac{3\lambda_1^2v^2}{64\pi^2}\left(-\frac{2}{3}+\frac{2\pi}{3\sqrt{3}}\right)+\frac{3\lambda_3^2v^2}{64\pi^2}\left(\frac{2}{3}-\frac{2\lambda_1}{3\lambda_3}\right)\nn\\
  &\phantom{=}\quad +\frac{22N^2 y_1^2}{16\pi^2}\Bigg[-2\left(y_1^2v^2-\frac{\lambda_1v^2}{12}\right)\nn\\
  &\hspace{0.7in}-3\int_0^1\text{d}x\left(y_1^2v^2-\frac{x(1-x)}{2}\lambda_1v^2\right)\ln\left(1-x(1-x)\frac{\lambda_1}{2y_1^2}\right)\Bigg],\nn\\
  &\simeq \frac{\lambda_3^2-88N^2
    y_1^4}{32\pi^2}v^2,\nn\\
&=\frac{\varepsilon}{32\pi^2}v^2,\nn\\
  M_{\phi_2}^2 &= \frac{\lambda_3v^2}{2}+\frac{\lambda_1\lambda_3v^2}{64\pi^2}\left(\ln\frac{\lambda_1v^2}{2{\mu}^2}-1\right)+\frac{\lambda_2\lambda_3v^2}{64\pi^2}\left(\ln\frac{\lambda_3v^2}{2{\mu}^2}-1\right)\nn\\
  &\phantom{=}\quad+\frac{\lambda_3^2v^2}{16\pi^2}\left(\ln\frac{\lambda_3v^2}{2{\mu}^2}+\int_0^1\text{d}x\ln\left(x^2+(1-x)\frac{\lambda_1}{\lambda_3}\right)\right)\nn\\
  &\phantom{=}\quad +\frac{22N^2 y_2^2}{16\pi^2}\Bigg[y_1^2v^2-\frac{\lambda_3v^2}{12}-3\left(y_1^2v^2-\frac{\lambda_3v^2}{12}\right)\left(\ln\frac{y_1^2v^2}{{\mu}^2}\right)\nn\\
  &\hspace{0.7in}-3\int_0^1\text{d}x\left(y_1^2v^2-\frac{x(1-x)}{2}\lambda_3v^2\right)\ln\left(1-x(1-x)\frac{\lambda_3}{2y_1^2}\right)\Bigg]\label{eq:phi2polemass},\\
  &\simeq \frac{\lambda_3v^2}{2}+\frac{\lambda_2\lambda_3v^2}{64\pi^2}\left(\ln\frac{\lambda_3v^2}{2{\mu}^2}-1\right)+\frac{\lambda_3^2v^2}{16\pi^2}\left(\ln\frac{\lambda_3v^2}{2{\mu}^2}-2\right)\nn\\
  &\phantom{=}\quad+\frac{22N^2 y_2^2}{16\pi^2}\Bigg[y_1^2v^2-\frac{\lambda_3v^2}{12}-3\left(y_1^2v^2-\frac{\lambda_3v^2}{12}\right)\left(\ln\frac{y_1^2v^2}{{\mu}^2}\right)\nn\\
  &\hspace{0.7in}-3\int_0^1\text{d}x\left(y_1^2v^2-\frac{x(1-x)}{2}\lambda_3v^2\right)\ln\left(1-x(1-x)\frac{\lambda_3}{2y_1^2}\right)\Bigg].\nn
\end{align}
The first lines of Eqs.~\eqref{eq:phi1polemass}
and~\eqref{eq:phi2polemass} are the complete one-loop expressions for
the pole masses, while the second line on Eq.~\eqref{eq:phi1polemass}
uses Eq.~\eqref{eq:vev} and shows that the whole expression is of one-loop
order and that it has no explicit $\mu$ dependence. The last line in
both equations is further simplified using the approximation valid at $\mu_0$
that $\lambda_1$ is small.  Observe that these scalar masses differ
from the curvature of the effective potential at the minimum.  This is
because the effective potential is computed at zero external
momentum,while the pole mass is computed at a momentum equal to the
pole mass itself.

It is instructive to check that these masses are RG-invariant. The
important observation is that the vacuum expectation value, $v$,
transforms under the RGE with the anomalous dimension of $\phi_1$:
\begin{equation}
\label{eq:vevrun}
  \frac{\partial v}{\partial t} = \gamma_{\phi_1}v = -\frac{11N^2y_1^2}{16\pi^2}v\,.
\end{equation}
Using this, the  above expressions for the pole masses and  the beta functions in
Eq.~\eqref{eq:betafunctions}, one can verify that 
\begin{equation}
	\frac{\partial M_\psi}{\partial t}=\frac{\partial M_\chi}{\partial t}=\frac{\partial M_{\phi_1}}{\partial t}=\frac{\partial M_{\phi_2}}{\partial t}=0\,,
\end{equation} 
up to terms of order of two loops.  This is of course expected, but the
explicit computation gives  a check of the above expressions. For
this check we have not used the approximation that $\lambda_1$ is
small.  This approximation is only valid for $\mu\sim\mu_0$, but we
will be examining shortly RG-trajectories that extend to $\mu\ll\mu_0$
where the approximation breaks down.


\section{Dilaton}

\subsection{Dilatation Current}
The dilatation current, $\mathcal{D}^\mu$ is related to the improved
stress-energy tensor through $\mathcal{D}^\mu =
x_\nu\Theta^{\mu\nu}$~\cite{Callan:1970ze}.  There are two important
properties of the improved energy momentum tensor. First, it is not
renormalized, so it has no anomalous dimensions. And second, it is
such that the divergence of the dilatation current is just the trace
of the stress-energy tensor,
$\partial_\mu\mathcal{D}^\mu=\Theta^\mu_\mu$. A simple way of
computing this tensor is by re-writing the model in a general
covariant fashion, with a background metric $g_{\mu\nu}$, taking
$\Theta^{\mu\nu}=-2\frac{\delta}{\delta g_{\mu\nu}}S_m$ where $S_m$ is
the action integral (exclusive of the Hilbert-Einstein term) and then
re-setting the metric to the trivial one $g_{\mu\nu}=\eta_{\mu\nu}$.  From
the Lagrangian in~(\ref{eq:model}) we have
\begin{multline}
\label{eq:thetamunu}
  \Theta^{\mu\nu} = -F^{a\mu\lambda}F^{a\nu}_\lambda+\frac12\bar\chi i(\gamma^\mu D^\nu+\gamma^\nu D^\mu)\chi+\frac12\bar\psi i(\gamma^\mu D^\nu+\gamma^\nu D^\mu)\psi\\
   +\partial^\mu\phi_i\partial^\nu\phi_i
  -\frac12\kappa(\partial^\mu\partial^\nu-g^{\mu\nu}\partial^2)\phi_i^2
  -g^{\mu\nu}\mathcal{L}\,.
\end{multline}
The term proportional to $\kappa$ is the improvement: it is
automatically conserved and is itself a total derivative so its
integral vanishes, leaving the generators of energy and momentum
$\int d^3x\; \Theta^{0\mu}$ unmodified. The improved tensor
corresponds to setting $\kappa=1/3$.  

Classically the trace of this tensor vanishes and therefore the
divergence of the dilatation current vanishes too. The theory is
classically scale invariant. As is famously known this is no longer
the case once quantum effects are included. Instead one has a ``trace
anomaly:''~\cite{Coleman:1970je,Adler:1976zt}
\begin{equation}
  \label{eq:traceanomaly}
  \Theta^\mu_\mu = \gamma_{\phi_1} \phi_1\partial^2\phi_1 + (4\gamma_{\phi_1}\lambda_1-\beta_{\lambda_1})\frac{\phi_1^4}{24}+\ldots,
\end{equation}
where we have kept only the terms involving $\phi_1$ since these will
play a role in our discussion below. The terms involving the field
anomalous dimension $\gamma_{\phi_1}$ are often overlooked. They can
be ignored when application of the equations of motion is valid but
may play a role in off-shell matrix elements or Green
functions.\footnote{There is an interesting technical subtlety
  here. The equations of motion that can and should be used are those
  for the bare fields~\cite{Politzer:1980me}. The use of the equation
  of motion in Eq.~(\ref{eq:traceanomaly}) gives that the terms
  proportional to $\gamma_{\phi_1}$ cancel.  On the other hand, the
  insertion of the anomaly into a matrix element would have us replace
  $-M_{\phi_1}^2$ for $\partial^2$ but since this mass starts only at
  one-loop order its product with $\gamma_{\phi_1}$ would give a higher order
  effect and spoil the cancellation against the rest of the
  $\gamma_{\phi_1}$ terms. We have verified by explicit computation that in fact the
  cancellation is not spoiled. To this end one must use the relation
  in Eq.~\eqref{eq:vev} that effectively trades $\lambda_1$ for one-loop
  terms.}
There is a simple indirect indication that these additional
terms must be included: since $\Theta^{\mu\nu}$ is not renormalized
the trace anomaly must be an RG-invariant, and the
$\gamma_{\phi_1}$-terms are required for this
purpose~\cite{Grinstein:1988wz,*Robertson:1991jk}.

\subsection{Dilaton}
As a pseudo-Nambu-Goldstone boson the dilaton state $|\sigma\rangle$ should be created
by acting on the  vacuum with the spontaneously broken dilatation
current. In analogy with PCAC we define a dilaton decay constant
$f_\sigma$  and a dilaton mass $M_\sigma$ so that 
\begin{equation}
\label{eq:sigma-trace}
  \langle0|\partial_\mu\mathcal{D}^\mu|\sigma\rangle =
  \langle0|\Theta^\mu_\mu|\sigma\rangle_{x=0} = -f_\sigma M_\sigma^2\,. 
\end{equation} 
This equation contains a particular combination of decay constant and
mass and we would like to be able to distinguish between them. The
matrix element of the current itself (which in PCAC gives the decay
constant directly) is not very useful because of the
explicit coordinate dependence. Instead consider the energy momentum
tensor, before taking the trace:
\begin{equation}
\label{eq:sigma-untrace}
  \langle0|\Theta^{\mu\nu}(x)|\sigma\rangle = \frac{f_\sigma}{3}\left(p^\mu
    p^\nu-g^{\mu\nu}p^2\right)e^{ip\cdot x}
\end{equation} 
The form of this equation is fixed by conservation of the
stress-energy tensor and that its trace
is given by Eq.~\eqref{eq:sigma-trace}. Note that in
Eq.~\eqref{eq:sigma-untrace} the  momentum is on-shell,
$p^2=M_\sigma^2$. 

In order to compute $f_\sigma$ and $M_\sigma$  we must first identify a state
in the spectrum of our model as the dilaton.  Were we in the exact
symmetry limit there would be a unique one-particle state that couples
to the stress energy tensor, making the identification of  the dilaton
straightforward. If the symmetry is not exact but approximate we
expect the dilaton to be a spinless state that (1) couples most strongly to the stress energy
tensor and (2) is the lightest state that does. It is easy to see that
the state of mass $M_{\phi_1}$ fits the bill. First, it is the
lightest of the two
spinless one-particle states in the spectrum, which is clear since the
perturbative expansion for its mass starts at one-loop order. To see
that it couples more strongly, note that when expanding the fields about the vacuum  $\vev{\phi_1}=v$ and
$\vev{\phi_2}=0$  in the stress
energy tensor, the only field that appears linearly is
$\phi_1$. Therefore  the only one-particle state that has tree level
overlap with the stress energy tensor is the state created by
$\phi_1$. 

With this identification we can now compute the decay constant to tree
level. Shifting the fields in Eq.~\eqref{eq:thetamunu}  and
concentrating on terms that can give $p^\mu p^\nu$ in the matrix
element, we have
$\Theta^{\mu\nu}=-1/3v\partial^\mu\partial^\nu\phi_1+\cdots$. The
ellipsis stand for terms that contribute only at  higher order than tree
level. Hence we read off $f_\sigma=v$. And, of course,
$M_\sigma=M_{\phi_1}$.

The anomaly equation gives us  a non-trivial check of this
identification. Going to shifted fields in  the anomaly Eq.~(\ref{eq:traceanomaly}), we have
\begin{equation}
  \Theta^\mu_\mu =\gamma_{\phi_1} v\partial^2\phi_1 +
  (4\gamma_{\phi_1}\lambda_1-\beta_{\lambda_1})\frac{v^3\phi_1}{6}+\ldots
\end{equation}
Taking the matrix element of this, working to lowest order (tree level
in the graphs). we obtain
\begin{equation}
  \langle0|\Theta^\mu_\mu|\sigma\rangle_{x=0} =
  -\gamma_{\phi_1}vp^2-\frac{\lambda_1^2+\lambda_3^2 - 88N^2
    y_1^4}{32\pi^2}v^3+\ldots .
\end{equation}
This agrees with Eq.~\eqref{eq:sigma-trace} if we use our identifications
\begin{equation}
f_\sigma=v\qquad\qquad\text{and}\qquad\qquad  
M_\sigma^2 = M_{\phi_1}^2=\frac{\varepsilon}{32\pi^2}v^2.
	\label{eq:dilatonmass}
\end{equation}
We have dropped the $\gamma_{\phi_1}vM_\sigma^2$ and $\lambda_1^2v^3$ terms for consistency.

Since the improved stress energy tensor is not renormalized the decay
constant $f_\sigma$ must be an RG-invariant quantity.  $M_\sigma$ is
also RG-invariant as any physical mass must. The expressions we have
found are not RG-invariant only because we have expressed them in
lowest order of perturbation theory. The pole mass, which we have
already discussed earlier, is explicitly seen to be RG-invariant to
one-loop order for the trivial reason that it itself starts at one-loop
order. On the other hand, the vacuum expectation value runs like the
field, Eq.~\eqref{eq:vevrun}. If $Z(t)$ is the wave-function
renormalization factor, $\partial Z/\partial t=2\gamma_{\phi_1}Z$,
$Z(0)=1$, where $t=\ln(\mu/\mu_0)$, then $f_\sigma=v/Z^{1/2}$ is an
RG-invariant, the RG-improved version of the previous result.


\section{Phase Structure}
\label{sec:phase}
We return here to the study of the phase structure of the model, posed
earlier in Sec.~\ref{sec:vac}. Let us recapitulate from there. A
perturbative study of the vacuum structure of the theory requires that
we limit our attention to a region of parameter space where
$\lambda_1$ is small.  Then the model possesses a new, non-trivial
minimum provided~\eqref{eq:posEig} is satisfied. Neither of these
conditions are satisfied in the neighborhood of the IR-fixed point.
However, we can take any point in the vicinity of the fixed point and
ask whether its RG-trajectory maps back at some large RG-time $t$ to
the region where a perturbative analysis of the effective potential is
valid and gives a non-trivial minimum. In fact, by reversing the
process, that is, by starting with a well chosen point at large
RG-time $t$ and then running towards the IR, we argued that there
always exist points arbitrarily near the IR-fixed point for which the
symmetry is spontaneously broken.  Choose  coupling constants at
some renormalization scale $\mu_0$  that give a non-trivial minimum
and so that the expectation value is small $\vev{\phi_1}\ll
\mu_0$. The coupling constants will run as in the mass independent
scheme towards the IR-fixed point and will get closer the smaller the
value of $\vev{\phi_1}$.  At $\mu\sim\vev{\phi_1}$ the running will be
modified and the trajectory will not hit the fixed point, but will
have gotten very close.

Now let's complete the picture. When $\mu$ becomes of the order of  the
physical mass of the heaviest particles in the spectrum the running of
the couplings is modified. For $\mu$ below the scale of that mass the
beta function becomes effectively the one for the model in the absence
of those massive particles, that is, the heavy particles are ``integrated
out.'' As $\mu$  is further decreased one sequentially integrates out
all massive particles in the model. This all occurs near the fixed
point so all couplings are still perturbative, but now all scalars and
spinors are integrated out. The Yukawa and self-couplings stopped
running and become uninteresting since  the effective theory contains
only massless Yang-Mills vectors. Now the beta function of this
effective theory is very much like that of QCD: the coupling constant
quickly runs to strong coupling, 
\begin{equation}
g^2(\mu)\approx\frac{g_*^2}{1+\frac{g_*^2}{16\pi^2}\frac{22N}{3}\ln\frac{\mu}{\vev{\phi_1}}}
\end{equation}
The spectrum of the effective theory is that of a theory of pure glue,
that is glueballs, of mass
\begin{equation}
M_g\sim \vev{\phi_1} e^{-\frac{3}{22N}\frac{16\pi^2}{g_*^2}}
=\vev{\phi_1} e^{-{225}/{44\delta}}
\end{equation}
So the spectrum of the model consists of two massive scalars and $n_f$
massive fermions with masses given in Sec.~\ref{sec:spectrum} plus
glueballs with masses $M_g$. The lighter scalar can be identified with
the dilaton and its mass is given by Eq.~\eqref{eq:dilatonmass}.

We can repeat the analysis, only now starting from a set of coupling
constants that does not satisfy the condition~\eqref{eq:posEig} at
$\mu_0$.  The potential now remains positive up to large values of
$\phi_1/\mu_0$ and one expects that by the time it starts decreasing
perturbation theory ceases to be applicable. So we expect the true
vacuum is at the origin of field space
$\vev{\phi_1}=\vev{\phi_2}=0$. There is no spontaneous scaling
symmetry breaking, all particles are massless. As $t\to-\infty$ the
RG-trajectories run into the IR-fixed point. 

The following picture emerges: the theory has two phases. The
parameter space of the model, which we identify with the space of
couplings at a fixed renormalization scale $\mu_0$, is split in two
regions. In region I the spectrum is massless and all RG-trajectories
run into the IR-fixed point. In region II there are no massless
particles and RG-trajectories do not end at the IR-fixed point. There
is a boundary between these phases, a hypersurface in the parameter
space of the model. The fixed point lies on this surface.  

The expectation value $\vev{\phi_1}$ vanishes in region I, but does
not in region II. The transition is discontinuous: by dimensional
transmutation, there is a non-trivial minimum of $V_{\text{eff}}$ at
an arbitrary\footnote{Arbitrary, but not extreme: the logs of
  $\vev{\phi_1}/\mu_0$ cannot be too large if the perturbative
  analysis is to remain valid.} value of $\vev{\phi_1}$ provided
$\lambda_3^2 - 88N^2 y_1^4$ is positive, no matter how small. Since
the physical content is preserved by flows we see that the surface
itself is RG-invariant.

But perhaps we have rushed into conclusions.  Firstly,
when~\eqref{eq:posEig} is not satisfied the effective potential is
unbounded from below as one moves along the $\phi_1$ axis towards
large values of $\phi_1$. We stated without justification that at
large $\phi_1$ perturbation theory breaks down and one expects the
potential stays bounded from below. But there is no guarantee of this,
and even if the potential stays bounded it may develop a new global
minimum at large $\phi_1$.  Perhaps none of region I is physical?
And secondly, in order to reach the vicinity of the IR point, which
is AB's prescription for obtaining a light dilaton, we argued we
can choose $\vev{\phi_1}$ small enough that our RG-trajectory will get
there. But how do we know that this does not occur only for such small
$\vev{\phi_1}$ that the logs in the effective potential become too large,
again invalidating the analysis?

Fortunately we can go a long way towards settling these issues by
explicit computation.  Inasmuch as the potential becomes one
dimensional (the minimum or the unbounded direction both lie on the
axis) we can use the RGE to re-sum the leading logs hence extending
the region of validity of the computation to the whole space of
perturbative parameters. For the effectively one dimensional case the
effective potential is $V_{\text{eff}}=\frac1{24}\bar
\lambda_1(t,\lambda_1)Z(t)^2\phi_1^4$~\cite{Duncan:1985ab}. Here $t=\ln(\phi_1/\mu_0)$, $Z$
is a wave-function renormalization factor and $\bar\lambda_1(t,\lambda_1)$
is the running coupling constant, defined with boundary condition
$\bar\lambda_1(0,\lambda_1)=\lambda_1$. The first objection above is
settled as follows: for any RG-trajectory for which
$\bar\lambda_1$ stays positive we can assert the minimum of
$V_{\text{eff}}$ is at the origin of field space and there is no
symmetry breaking. The only caveat is that we cannot trust the
calculation at very large $t$ where  the scalar couplings
become non-perturbatively large. Recall the model has Landau poles so
it either is considered as a cut-off model or as the low energy limit
of a complete theory. 

The second objection can also be settled by following the trajectory
towards the IR. If at any point along the trajectory the running
coupling turns negative then there will be a minimum away from the
origin in field space, symmetry will be broken and a pseudo
Nambu-Goldstone boson associated with the breaking of scale invariance
will appear in the spectrum. One can then follow the trajectory and
determine how close it gets to the IR-fixed point. This is somewhat
unnecessary, since  we already established in the previous two
sections that for small $\varepsilon$ we get a light dilaton. 

Although the model is perturbative, we do not know how to analytically
integrate the RG trajectories. But it is quite straightforward to
investigate them numerically.  It is beyond the scope of this work to
conduct an exhaustive study of the phase diagram numerically. Instead
we follow the trajectories from some initial points at $\mu_0$ to gain
confidence the picture we have painted is not obviously flawed.  We
use $N=20$, $n_f=11N/2$, $\delta=0.2$.  First we take $ g(\mu_0) =
\frac{4}{9}g_\ast$, $y_1(\mu_0) = 0.45y_{1\ast}$, $y_2(\mu_0) =
\frac{1}{5}y_{2\ast}$, $\lambda_1(\mu_0) =
\frac{1}{30}\lambda_{1\ast}$, $\lambda_{2}(\mu_0) =
3\lambda_{2\ast}$, $\lambda_{3}(\mu_0) =5.2\lambda_{3\ast}$..
This set of parameters does not satisfy~\eqref{eq:posEig}.  The
effective potential doesn't develop a non-trivial minimum,the running
coupling $\bar \lambda_1$ remains positive. The theory flows to the IR
fixed point.  Next we analyze the case when $ g(\mu_0) =
\frac{4}{9}g_\ast$, $y_1(\mu_0) = 0.32y_{1\ast}$, $y_2(\mu_0) =
\frac{1}{5}y_{2\ast}$, $\lambda_1(\mu_0) =
\frac{1}{30}\lambda_{1\ast}$, $\lambda_{2}(\mu_0) =
3\lambda_{2\ast}$, $\lambda_{3}(\mu_0) = 5.2\lambda_{3\ast}$.
Naively, this theory seems to flow to the IR-fixed point as well.  But
in this case, the effective potential does develop a minimum at
$\ln\left({v^2}/{\mu_0^2}\right) \simeq -58$.  We estimate the
fractional correction to the effective potential from higher orders in
the loop expansion to be of order
\begin{equation*}
  \left|\frac{Ng^2}{16\pi^2}\ln\left(\frac{y_1^2v^2}{\mu^2}\right)\right|\simeq 0.2
\end{equation*}
Thus we can trust the minimum we find using perturbation theory.  With
this vev, the spectrum is $ {M_{\psi,\chi}}/{v} \simeq
8.5\times10^{-3}, {M_{\phi_1}}/{v} \simeq 7.9\times10^{-4},
{M_{\phi_2}}/{v} \simeq 9.5\times10^{-2}$. The scale $\mu_0$ is some
13 orders of magnitude larger than the vacuum expectation value $v$,
but it is unphysical.

We have studied numerically the transition between these two parameters
sets by varying $y_1(\mu_0)$ or $\varepsilon(\mu_0)$.  When $y_1(\mu_0)$ is
sufficiently large, or when $\varepsilon$ turns negative, we change from a
broken phase to the symmetric phase as expected from
Eq.~\eqref{eq:potentialatvev}. Note that with our particular value of
parameters, the theory is close to the  boundary of the
broken/symmetric phases.

\subsection{Relevant Perturbations}
Suppose we consider a modification of the model, one in which scale
invariance is explicitly broken. This is accomplished by adding
relevant perturbations. If the symmetries of the model are to be
preserved only mass terms can be added. This enlarges the parameter
space of the model. The origin of all the relevant-perturbation axes 
corresponds to the parameter space described in the previous paragraphs,
and it is on that hyperplane that the IR-fixed point lies together
with the two phases and the hypersurface separating them. 

Far away from this hyperplane, a long ways along the
relevant-perturbation axes, the physics is very simple: scalars and
spinors have hard masses and below the scale of those masses they
decouple so as to leave only light glueballs in the spectrum. A more
interesting region of parameter space is the direction of large scalar
masses and small spinor masses. Then the scalars decouple and one is
left with a  Banks-Zaks-like model. Only it  does not run into an
IR-fixed point because the spinors eventually decouple, the
YM-coupling then runs strongly and glueballs form. Only at zero spinor
mass do we see that  our IR-fixed point
is really part of an IR-fixed  hyperline. 

What is the fate of the two phases as one extends into the new axes?
In the symmetric phase the addition of hard masses can only make the
vacuum at the origin of field space more stable. The spectrum is
modified, particles are massive now and there is no IR-fixed point
(save for the zero spinor mass case). 

Analysis of the broken symmetry phase is more subtle. Provided we
stay very close to the origin of the new axes, so that the added mass
terms are really small perturbations, much smaller than the masses
obtained in the absence of the perturbations, then nothing changes
qualitatively and the quantitative changes to the spectrum are
small. As the strength of the relevant perturbations increase the
model may remain in a broken phase, depending on the precise nature of
the perturbations. But for large enough perturbations the dilaton will
be unrecognizable as a pseudo Nambu-Goldstone boson.

Summarizing, the two phase diagram does extend into the larger
parameter space. The fixed point becomes a  (hyper) line of fixed
points. For large perturbations the dilaton is gone.

\section{Discussion, Conclusion and Open Questions}
We have presented a model with an IR-fixed point, and demonstrated
that the model has two phases.  In phase I RG-trajectories run into
the IR-fixed point (in infinite RG-time). The scale symmetry is
approximate and explicitly realized and it becomes exact at the fixed
point.  In phase II scale symmetry is spontaneously broken. Of course,
scale invariance is also explicitly broken by the trace anomaly. The
trajectories don't reach the IR-fixed point but some get very close
and for those the explicit, relative to spontaneous, breaking of scale
invariance is small: A light dilaton appears in the spectrum. 

Analytic evidence for this picture was presented at length but 
the numerical support was scant. This is clearly an interesting
direction for future work. In particular, one could determine the
actual location of the phase transition.  Another direction for future
work is to find generalizations of the model. We do not know how
general this picture is or how difficult it may be to come about
models that display arbitrarily light dilatons (we were not aware of
any example prior to this work). 

Among new models one may try to construct some with the Standard Model
of electroweak interactions embedded in it. One could then test
whether the setup in Ref.~\cite{Goldberger:2008zz} works as
advertised. The authors there considered the possibility that the
standard model is embedded in an almost conformal, possibly strongly
interacting field theory with spontaneously broken scale invariance.
In the context of 4-dimensional strongly interacting near-CFTs obtained as
AdS/CFT-like duals of 5-dimensional non-factorizable geometries (RS
models) one encounters often the schematic Lagrangian describing the
dynamics:
\begin{equation}
\label{eq:CFT-perts}
\mathcal{L}=\mathcal{L}_{\text{CFT}} +\sum_n \lambda_n\mathcal{O}_n\,.
\end{equation}
The first term is a CFT while the sum that follows is an attempt to
capture the deviations (``deformations'') from the CFT by adding small
perturbations ~\cite{ArkaniHamed:2000ds,Rattazzi:2000hs}.  Obviously
this basic setup applies to our model, and because it is fully
perturbative model one should be able to verify the validity of some
general assertions. The deviations from conformality can be small in
one of two ways, either the anomalous dimensions $\gamma_n$ or the
coefficients $\lambda_n$ of the operators $\mathcal{O}_n$ are small.
On general grounds one can show that for $|\gamma_n|\ll1$ the effective
potential for the field $\chi$ whose expectation value gives rise to
the dilaton is~\cite{Goldberger:2008zz} 
\begin{equation}
\label{eq:GGS-v1}
V_{\text{eff}}(\chi)=\frac{M_\sigma^2}{4f_\sigma^2}\chi^4
\left[\ln\left(\frac{\chi}{f_\sigma}\right)-\frac14\right]+
\mathcal{O}(\gamma^2)\,.
\end{equation}
The case $|\lambda_n|\ll1$ is more cumbersome. Only in the case that
only one perturbation is added does one obtain a parameter-free
effective potential
\[
V_{\text{eff}}(\chi)=\frac{M_\sigma^2}{f_\sigma^2\gamma}\chi^4
\left[\frac1{4+\gamma}\left(\frac{\chi}{f_\sigma}\right)^\gamma-\frac14\right]+
\mathcal{O}(\lambda^2)\,,
\]
while for more than one perturbation occur one has the less restricted 
\[
V_{\text{eff}}(\chi)=\frac{M_\sigma^2}{f_\sigma^2}\chi^4 \sum_n
\left\{x_n\left[\frac1{4+\gamma_n}\left(\frac{\chi}{f_\sigma}\right)^{\gamma_n}-\frac14\right]\right\}+
\mathcal{O}(\lambda^2)\,,
\]
where the coupling constants have been traded for constants $x_n$ that
are constrained by $\sum_n \gamma_n x_n =1$. 

Any model with a conformal fixed point $g_*$ can be written in the fashion
of Eq.~\eqref{eq:CFT-perts}
\[
\mathcal{L}(g)=\mathcal{L}(g_*) + \left(\mathcal{L}(g)-\mathcal{L}(g_*)\right)
\]
where $g$ are coupling constants at arbitrary values. If $g$ is
sufficiently close to $g_*$ one is in the case $|\lambda_n|\ll1$
above, while if the region of couplings that includes $g$ and $g_*$ is
perturbative one expects $|\gamma_n|\ll1$. We need, in addition, that
the model display spontaneous breaking of scale invariance in the
vicinity of the fixed point. Our model furnished an explicit
example. The analogue of $\chi$ is our field $\phi_1$. Because it is
perturbative one has $|\gamma_n|\ll1$. Reassuringly, when the tree
level term in the effective potential of
Eq.~\eqref{eq:effectivepotential} is eliminated by use of
Eq.~\eqref{eq:vev}, and the expressions for dilaton mass and decay
constant in Eq.~\eqref{eq:dilatonmass} the resulting potential is {\it
  exactly} of the form of Eq.~\eqref{eq:GGS-v1}. To emphasize, the
dependence on the many coupling constants of our model is completely
contained now in only two parameters:  $M_\sigma$ and $f_\sigma$.

Finally, we  address one of the central questions we set out to
investigate: Is the AB estimate of the dilaton mass in walking
technicolor scenarios correct?  For AB, the dilaton mass is given by
\begin{equation}
  M^2_\sigma \simeq \frac{s(\alpha_\ast-\alpha_c)}{\alpha_c}\Lambda^2 \simeq \frac{N_f^c-N_f}{N_f^c}\Lambda^2,
	\label{eq:ABformula}
\end{equation}
where $\alpha_\ast$ is the coupling at the fixed point, $N_f$ is the
number of flavors and $\Lambda$ is the scale of chiral symmetry
breaking which occurs only if the critical coupling $\alpha_c$ is
below the fixed point, $\alpha_c<\alpha_\ast$, which in turn
corresponds to the number of flavors below a critical value,
$N_f^{c}$.  The middle expression in Eq.~\eqref{eq:ABformula},
relating the mass to the distance between the critical coupling and
the fixed-point, does not carry over to our model.  In our case, the
role of the critical value of the coupling constant $\alpha_c$ is
played by a critical surface, $\varepsilon = 0$, separating the
symmetric and broken phases.  But the mass of the dilaton is not
proportional to the distance between this surface and the fixed point
(however one defines distance): the fixed-point lies on the critical
surface and the dilaton mass vanishes everywhere on the surface.  The
rightmost expression in Eq.~\eqref{eq:ABformula}, however, has a
counterpart in our model.  In that formula $(N_f^c-N_f)/N_f^c$
measures how far the theory is from the critical point.  In our model
$\varepsilon$ plays the role of this quantity.  It measures how far
the theory is from the critical surface.  Moreover, both
$(N_f^c-N_f)/N_f^c$ and $\varepsilon$ can be made arbitrarily small
which in turn make the dilaton arbitrarily light compared to the
scale of symmetry breaking.  To the extent that one can arrange for
arbitrarily small $(N_f^c-N_f)/N_f^c$, AB's estimate of a
parametrically small dilaton mass is consistent with our analysis.

\begin{acknowledgements}
  We thank F. Sannino for bringing to our attention
  Ref.~\cite{Dietrich:2005jn} which arrives at a similar conclusion to
  that of~\cite{Appelquist:2010gy}.  We also thank L.~Vecchi for bringing
  to our attention Ref.~\cite{Vecchi:2010dd, *Vecchi:2010aj} which,
  via a different method, also arrives at the same conclusion
  as~\cite{Hashimoto:2010nw}.  This work was supported in part by the
  US Department of Energy under contract DOE-FG03-97ER40546.
\end{acknowledgements}


\bibliography{TCdb}{}
\bibliographystyle{apsrev4-1}

\end{document}